# Controlled and Continuous Patterning of Organic and Inorganic Materials by Induced Nucleation in an Optical Tweezers

Basudev Roy[a], Manish Arya[b], Preethi Thomas[c], Julius Konstantin Jürgschat[c], Venkat Rao[d], Ayan Banerjee[a]*, Chilla Malla Reddy[b]*, and Soumyajit Roy[c]*

We demonstrate for the first time controlled patterning by inducing nucleation of material from a dense aqueous dispersion or solution in an optical tweezers. A hot spot is formed on a glass surface by the trapping laser due to which a water vapour bubble is formed causing Gibbs-Marangoni convection of 
10 material around the bubble. This results in accretion of material around the bubble, which eventually nucleates into a crystalline state of the material. The nucleation site, when moved by translating the microscope stage of the optical tweezers apparatus, forms a pattern. We have demonstrated the technique using exotic inorganic materials such as soft oxometalates, and organic materials such as glycine, paracetamol, and a fluorescent dye such as perylene. We have written patterns over lengths of nearly 1
15 mm at the rate of 1 Hz, with best resolution of about 1 μm. The technique has potential for a wide range of applications ranging from solution processed printable electronics to controlled catalysis.

## Introduction

Nature displays a profuse abundance of patterns.[1,2] Thus, in modern times, having learned from Nature and faced with ever 
20 receding space and ever exploding information, efficient patterning has gained immense importance among materials scientists.[3,4] Added to the above challenges is the need for the development of a sustainable environment. Thus, as the modern world paces towards miniaturization of storage systems and 
25 information storage devices, an immediate challenge to materials scientists is to design patterns in a controlled and fast fashion using eco-friendly techniques.

Patterning has its own scales: macro, meso, micro and nano.[1] There are many ways of writing micro-patterns, the common ones 
30 being photolithography,[5] soft lithography,[6] electron beam lithography,[7] maskless lithography, ion beam lithography and the more recently developed electric field controlled deposition.[8] Many of these require complex apparatus to fabricate patterns that are quite expensive and time consuming as well. However,
35 laser-induced nucleation has provided a new alternative to generate spatially and temporally controlled patterns using a simpler set-up. For example, non-photochemical laser-induced nucleation (NPLIN)[9] has been used to produce controlled crystal nucleation in agarose gels where pulses of visible or near infra-
40 red laser light are employed to induce modification of the free energy in pre-nucleating clusters thereby making them supercritical.[10] These supercritical clusters then deposit as crystals, and many such crystals are then grown to form spatially controlled patterns. Such laser induced nucleation has received
45 considerable attention owing to the ease of this methodology. Moreover there exists previous work to show the possibility of inducing confined photo-polymerization and solidification due to radiation pressure induced by the application of a laser.[11][For a recent review on the topic see also:[12]] Laser induced nucleation –
50 when employed alone – involves different photo-physical phenomena such as optical Kerr effect, trapping in laser fields, transient pressure enabling material accumulation, [12,13] (see also: [14]) or photo-chemical methods of irradiating metastable solutions to generate nucleation centres. They share a common drawback:
55 very high laser power requirement (of the order of GW/cm$^2$!). This drawback has been circumvented recently by exciting the local plasmon resonance of a gold thin film, which can create 'hot spots', thereby facilitating nucleation with a CW laser in a localized region to form single crystals.[12] Likewise, exploiting
60 Gibbs-Marangoni convection, controlled patterning of nano-particles on a gold surface has been achieved.[15] Complex patterns have been formed using gold quantum dots – however, the patterns are not continuous, and the time scales involved are in the order of a few seconds, with the method also requiring a
65 functionalized (gold) substrate. These patterning methods lead to further imminent questions: Can the hot spots be created by other absorption processes such as exciting LMCT (Ligand Metal Charge Transfer)[16] type transitions to use nanostructures for writing, and can the patterning rates be made still faster (milli-
70 seconds)? Such options could lead us to exploit the unique reactivity of transition metals (unlike gold) and thereby facilitate unusual and low-cost solution-based patterning. The known catalytic activity of such transition metal oxide structures can also be an added bonus.[16]

75 In this paper, we report a controlled and continuous optical patterning technique using bubble-assisted nucleation in an optical tweezers, in a green solvent: water (see Figure 1, which is a cartoon depicting our technique). The optical tweezers is built around a CW-NIR (λ=1064 nm) laser with variable power (0-100 mW) incident on the sample. We have formed patterns following controlled nucleation on three classes of materials: 1) a class of newly synthesized dispersed soft-oxometalate (from here onward abbreviated as SOM)[17] nanotubes having comparatively high absorbance at $\lambda$ = 1064 nm resulting from an LMCT type transition, 2) organic molecules such as glycine and paracetamol, and 3) organic molecules (such as paracetamol) and fluorescent dyes (such as perylene where the pattern can be illuminated under light) loaded on the SOMs to demonstrate assisted nucleation. We observe that hot spots, and subsequently, continuous patterns can indeed be formed using the SOMs, and at much lower powers than that reported in literature.[12-14] Hot spots as well as patterns are also formed using organic molecules which have negligible absorbance at the laser wavelength, thus requiring much higher laser power. However, when the molecules are anchored on the SOMs, we observe assisted nucleation exploiting the excitation of SOM core due to LMCT type transition when exposed to the intense trapping beam. The organic molecules have been chosen keeping in mind the presence of hydrogen bonding and coordination sites. This technique is easily controllable and fast (time scales less than 1 s presently achieved for writing 1 mm long line patterns) for any optical patterning scheme. We perform a detailed analysis of the factors affecting the technique using diverse material systems and different operating conditions so as to optimize the technique for wide applicability as a solution-based patterning process.

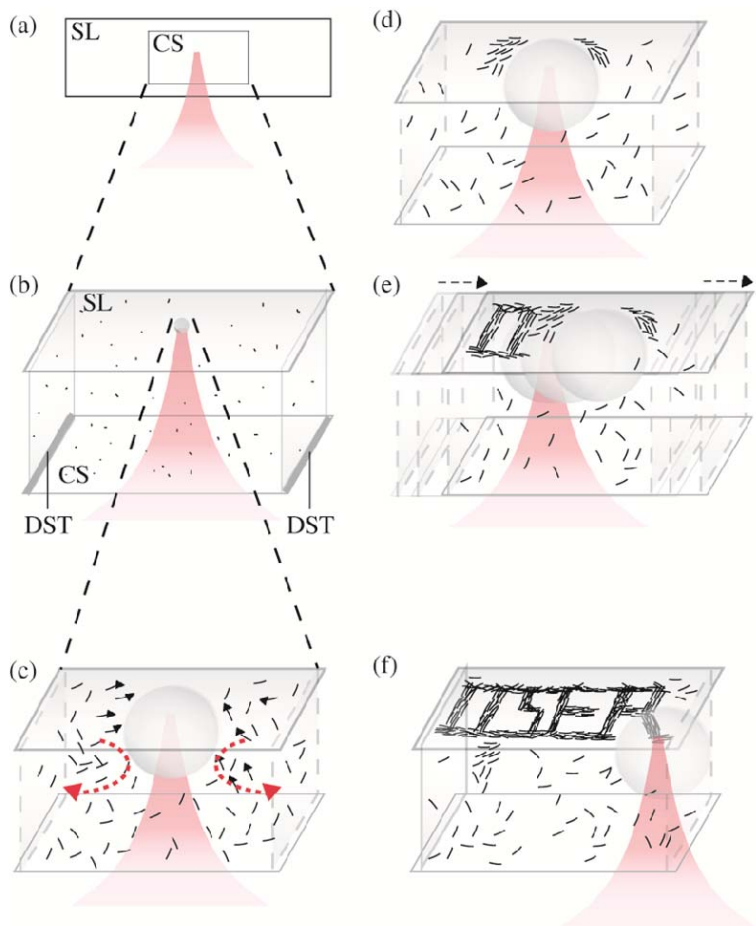

**Fig.1** A schematic of the technique used for patterning. (a) Sample holder consisting of a top microscope slide (SL) and bottom cover slip (CS) attached to each other by a double sided tape (DST shown in (b)). The sample holder contains an aqueous solution/dispersion of the material. The holder is placed on the automated precision scanning stage of an inverted microscope. A Gaussian beam (shown in red) is focused tightly near the top slide to form the optical tweezers. (b) The region around the cover slip zoomed in. The liquid has local density maxima of the dispersed soft material, where absorption from the focused laser beam is high – thus a hot spot is created on the slide giving rise to a water vapour bubble. A single bubble formed by the coalescence of several micro-bubbles is shown. (c) The region around the bubble zoomed in. The temperature gradient in the vicinity of the bubble gives rise to strong Gibbs-Marangoni convection currents shown in red dashed arrows. The circulating currents cause a flow of soft material (SOM nanotubes and/or other species) towards the bubble which is shown by black arrowheads attached to the indiviual particles. (d) The soft material gradually deposits at the base of the bubble (in the form of a ring which is not shown), and undergoes a phase transition to enter into a crystalline phase. (e) The sample chamber is then translated controllably by scanning the microscope scanning stage. The deposition of material at the base of the bubble ensures that the laser 'hot spot' moves concomitantly due to continuous absorption by the material, thus causing thermo-capillary migration of the bubble, a process which continuously deposits fresh material at the bubble base. A continuous track of crystalline (hard) material is therefore formed. (f) The microscope scanning stage can be moved horizontally and vertically to form a pattern of the word 'IISER' with the nucleating material using a single bubble.

## Experimental

### Optical System

The patterning is performed over a surface where a high density aqueous solution or dispersion of the material to be patterned is exposed to an optical tweezers that consists of a CW laser beam focused to a diffraction limited spot using a high numerical aperture objective lens.[18] The optical tweezers is constructed around an inverted microscope (Zeiss Axiovert. A1 Observer). A 100X, 1.4 N.A. oil immersion microscope objective (Zeiss plano-apochromat, infinity corrected) is used to couple 1064 nm light from a diode laser (Lasever LSR1064ML) into the sample holder to attain a spot size of waist radius ~500 nm. The laser power can be varied from 0 to 100 mW after the objective. The sample solution/dispersion is placed or sandwiched in a chamber consisting of a 1 mm thick standard glass slide (top surface) and a 160 μm thick glass cover-slip (bottom surface) with double sticking tape on the sides to control the spacing. About 75 μL of solution is inserted into the holder, which is then placed in the microscope scanning stage which is stepper motor controlled with a Ludl MAC5000 XY stage controller operated using a joystick. The total travel range is 130 x 100 mm with resolution of 100 nm. The process of crystallization is observed using a CCD camera (Axiovision) running at a best rate of 30 frames per second. Imaging is performed at the back focal port of the microscope using a combination of 10X/40X/100X objectives and external optics arranged outside the microscope.

### Synthetic Procedure

a. Molybdenum based Soft-Oxometalate: The SOM was prepared by dissolving 817.6 mg of ammonium heptamolybdate tetrahydrate (from Sigma Aldrich) in 4 mL of water which was warmed until a homogeneous dispersion was obtained that scattered light from a laser. This dispersion was then cooled to room temperature before being used for optical patterning experiments.
b. Glycine: The high density solution of glycine (from Sigma Aldrich) was made by dissolving 65 mg of glycine in 0.5 mL of water and sonicating it for 1 min.
c. Paracetamol: 5 mg of paracetamol (from Sigma Aldrich) was dissolved in 0.5 mL water and sonicated for 1 min.
d. SOM with Paracetamol: Ammonium heptamolybdate tetrahydrate (817.6 mg) was dissolved in 4 mL of water and was warmed until a homogeneous dispersion was obtained. This dispersion was then cooled to room temperature. Paracetamol (10 mg) was added to this dispersion in the mole ratio (1:10).

## Results and Discussions

### Choice of materials and solvent

Our choice of materials is driven by the necessity to form hot spots with minimum dependence on surface properties. Hence, absorption of the material could help in hot spot formation without the need for intricate surface modification. For this reason, SOMs were ideal for our present methodology. SOMs[17] are a class of materials that have a diffuse or soft interface and have colloidal length-scales and comprise of several thousands of oxometalate building blocks stitched by supramolecular interactions. Several such SOMs have been synthesized by us and others.[17] Some SOMs are spontaneously formed,[19-21] while other formations are directed, using colloidal structural directing agents.[22-23] In this work, we synthesize SOM nanotubes (molybdenum based) and form patterns using a dispersion of such tubes, and also use them individually as templates for loading organic materials (see later). They are characterized in detail by microscopic and spectroscopic techniques. To check whether this method works with organic molecules having poor absorptivity in the IR region, we used glycine and paracetamol solutions in water, and observed that hot spots, and thereby patterns could be formed in those cases as well, albeit by using almost five times higher power for the trapping laser. The power threshold for pattern formation was substantially reduced once we loaded organic molecules such as paracetamol, and a fluorescent dye, perylene, as a substrate on the SOMs. Note that all these organic molecules have hydrogen bonding and coordination sites available to bind with the SOMs.

We choose water as a solvent mainly due to the fact that transition metal oxides with LMCT bands and several other organic molecules with coordinating groups dissolve readily in water. SOMs, on the other hand, disperse in water. In addition, given the need for a cheap and green solvent, water becomes a natural choice. Hence the technique used here could be employed for cheap and eco-friendly electronic printing.

### Nucleation process

*Bubble nucleation:* For best results, the trapping laser is focused near the top slide. Regions of high material density in the liquid occurring in the proximity of the top slide get hotter than the surrounding media due to higher absorption of the focused laser beam to form hot spots on the surface. Such a region is probed by an initial scan of the microscope stage and a hot spot is located. Several water vapour micro-bubbles are formed in the vicinity of the hot spot in time scales of the order of microseconds. It is interesting to note that a similar hot spot is also formed when the laser is focused at the bottom surface (cover slip), but bubbles formed in that region would be pushed upwards immediately by buoyant forces so that nucleation would not occur at all. However, the micro-bubbles formed near the top slide immediately coalesce to form a large bubble following classical nucleation process,[24] and the single bubble grows in size very quickly (again in tens-hundreds of microseconds depending on the laser intensity). Additionally Gibbs-Marangoni convection sets in, leading to the accumulation of material towards the 'bubble base' – which is essentially the contact area of the bubble on the top slide – in the form of a ring.[15] There is also a direct evidence of bubble nucleation by Ostwald ripening which we monitor by measuring the growth of the bubble diameter with growth of a bubble, the trapping laser power was blinked or on/off modulated by an acousto optic modulator (AOM) for controlled time periods. We increased the laser on-time gradually

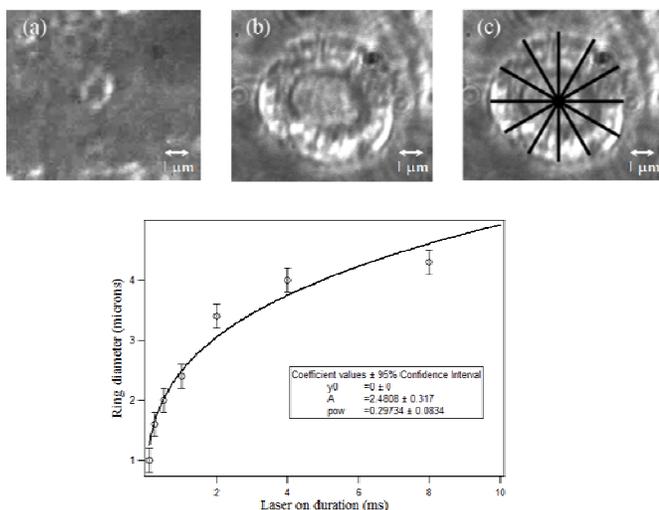
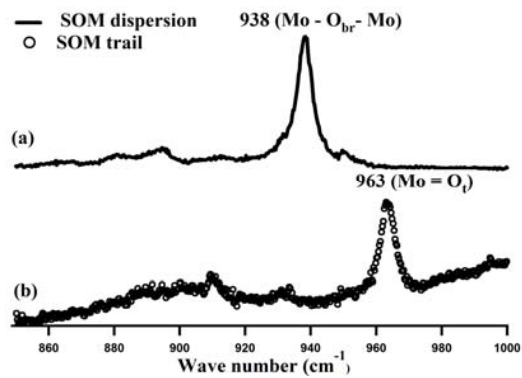

**Fig.2** (a) A ring of deposited SOMs, now in the crystalline state. The trapping laser has been turned on for 100 μs. The ring diameter is around 1 μm. (b) Ring of deposited SOMs with the laser turned on for 8 ms, the
5 diameter has now increased to around 4.5 μm. (c) Determination of the diameter of the ring formed by deposition of the nucleating material. Since the ring is not exactly annular, we measure the diameter at six locations covering the entire manifold and take an average. (d) Ostwald ripening in action: Plot of ring diameter vs laser on duration. The increase
10 of the diameter follows a power law behavior with time, with a fit exponent of around 0.30(8). This is clearly indicative of an Ostwald ripening process occurring in the nucleating material as the laser is kept on for longer periods of time.

15 and measured the size of the material ring deposited at the base of the bubble, and plotted the increase in ring diameter with increasing laser on-times. The ring diameter was measured by using reference polystyrene beads of known diameter stuck to the glass slide. The final diameter calculated was the average of six
20 different estimates at different locations along the entire manifold of the ring. (Figure 2c) The plot of ring diameter as a function of time is shown in Figure 2d. Note that the material ring is essentially the footprint of the bubble and would therefore carry the signatures of Ostwald ripening. A fit to the data (ring
25 diameter against time) yields a power law behaviour with an exponent close to 0.33 – which strongly demonstrates Ostwald ripening (Figure 2d).

*Crystal Nucleation:* It is interesting to note that when the laser is turned off, the material ring (viscous) formed at the bubble-
30 base undergoes a soft-to-crystalline phase transition that ultimately forms crystals.[25] This is consistent with the two stage crystal nucleation theory where the end of the first stage is manifested by a phase transition from a viscous soft state to a crystalline hard state of matter.[26-28] The crystallization is verified
35 using Raman spectroscopy (Figure 3). We use the SOMs to study this phase transition since they are most easily patterned. Clearly there are two steps in this process, viz., i) accumulation of the SOMs in dispersion (described previously) by the trapping laser and ii) their transition to crystalline oxometalates when the
40 trapping laser is turned off. The Raman spectra of the SOMs in dispersion have an intense Mo-$O_{br}$-Mo band at around 938 cm$^{-1}$, which is due to the breathing mode of the SOM nanotubes in dispersion (Figure 3a).[22] However on crystallization this band almost disappears whereas an intense Mo=$O_t$ band is observed as
45 in crystalline state these modes are active and numerous, which implies that crystallites of oxometalates are formed in the optical field (Figure 3b).

**Fig.3** Raman spectral patterns: (a) Raman spectra of SOMs in an
50 aqueous dispersion (solid line), and (b) Raman spectra of the trails in a bubble-induced trail (open circles). The Raman spectra of the SOMs in aqueous dispersion, have an intense Mo-$O_{br}$-Mo band at around 938 cm$^{-1}$, which is due to the breathing mode of the SOM nanotubes in dispersion.[23] However, on crystallization,
55 this band disappears whereas an intense Mo=$O_t$ band is observed at 963 nm,thus implying that crystallites of oxomolybdates are formed due to the trapping beam.

## Pattern formation

60
After the bubble grows to a certain size, the laser spot is moved to a new position of the sample by scanning the sample holder translation stage. Now, the hot spot is concomitantly moved, and the bubble follows the hot spot due to Gibbs-
65 Marangoni convection – a phenomenon known as thermocapillary bubble migration.[29] Thermocapillary bubble migration using a pre-coated substrate with absorbing material has been reported earlier[29] – however, in our method the need for pre-coating is eliminated, since material is continuously deposited
70 at the base of the bubble and slightly adjacent to it, so that there is always an absorbing media present on the surface of the slide to create a hot spot when the laser is shifted away slightly from the bubble. The bubble could thus be described to be indirectly trapped by the trapping laser beam, and follows the beam as it is
75 scanned at high rates [around 1 mm/s or even higher], continuously causing crystallization of material (in these cases SOMs and other organic molecules) in contiguous rings, which finally forms the desired pattern having the diameter of the rings. In other words, this is indeed the translation of a trapped
80 nucleation site which continuously leaves a trail of crystallized material behind. A chosen pattern is then formed by scanning the stage in a particular pre-determined manner (Figure 4). The factors affecting pattern formation are described later. Note also, that the process is much more simpler when one attempts to
85 extend a pre-existing pattern. A bubble is very easily formed when the trapping laser is focused to a pre-existing crystallized material track, because this acts as a seed to ensure faster formation of the nucleating site which can then be translated

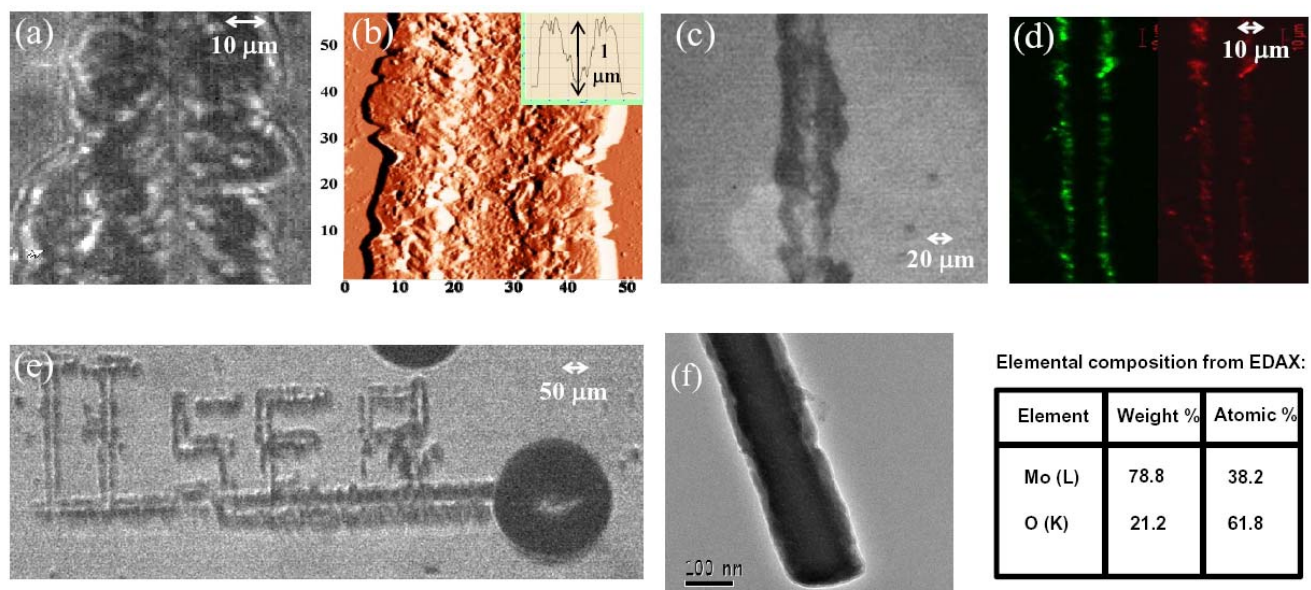

**Fig. 4** Pattern formation by nucleation: (a) Linear pattern of SOMs formed by controlled thermo-capillary migration of a bubble in a dispersion of SOM nanotubes. A continuous trail is formed having width of around 30 μm. (b) AFM image of the trail. It shows continuous deposition of material with higher accumulation at the sides which is understandable since the material deposition would be lesser near the center where the bubble is in contact with the top
5 slide. The trail width is around 40 μm. Inset shows the thickness of the trail to be 1μm. (c) Linear pattern formed with Glycine. Laser powers required are higher by around a factor of 5 compared to the SOMs to form a bubble and then to induce thermo-capillary migration by scanning the microscope stage. Material trail width is around 50 μm. (d) Linear pattern formed with the dye perylene loaded on the SOMs. Green fluorescence is observed from the material trail when irradiating with a laser at 488 nm, while red fluorescence is observed when the irradiating wavelength is modified to 561 nm. Trail width is around 15 μm. (e) Pattern 'IISER' written on the top slide with deposited SOM crystals using a single bubble that is shown at the lower right
10 corner of the picture (at the edge of the letter 'R'). The microscope scanning stage is translated horizontally and vertically in order to form the letters with the bubble following the stage movement continuously. Material is continuously crystallized in the wake of the bubble to form the pattern. (f) TEM images of a single SOM nanotube, with EDAX showing the elemental constituents (provided in tabular form) of the pattern IISER that was written. SAED pattern (inset) shows non-crystalline nature of the nanotubes.

efficiently to cause further crystallization of material in a
15 specified design.

For SOMs, the threshold for the bubble formation was found to be around 20 mW at the sample plane, which implied an intensity of only around 1.5 MW/cm$^2$ (more than an order of magnitude lesser than that reported in Ref 25) – a factor that can be
20 attributed to the high absorption levels observed in the electronic absorption spectra (EAS) of SOMs due to LMCT type transitions. A trail of deposited material of crystallized oxometalate is shown in Figure 4a. To verify the process of permanent deposition, AFM images were taken of the crystalline oxometalate patterns created
25 in the top glass slide. As seen in the inset of Figure 4b, the AFM image confirms deposition of material to a thickness of about 1 μm on the slide. Linear trails with deposited material were also obtained in organic molecules such as glycine and paracetamol using about five times higher laser power compared to the SOMs
30 to compensate for the low absorption of the former at the laser wavelength. A trail formed with glycine is shown in Figure 4c. However, when the same molecules are loaded on the SOMs, the power threshold for nucleation reduces drastically to the same value as that with only the SOMs. A trail of carboxylate
35 substituted perylene loaded on SOMs is shown in Figure 4d, which when irradiated with different laser wavelengths, namely 488 nm and 561 nm, produces green and red fluorescence respectively as is apparent from Figure 4d. Finally, we imprinted a pattern IISER on the slide by manually moving the microscope
40 translation stage in the required manner (Figure 4e). The resolution achieved was around 50 micrometers – note that this could be improved by several methods such as using smaller bubbles (employing lower laser intensity), or by reducing the laser spot size by employing either a higher NA microscope
45 objective, or by using a smaller laser wavelength. Figure 4f shows TEM images of a single SOM nanotube, with EDAX (Energy Dispersive X-ray Spectroscopy) showing the elemental constituents (provided in tabular form) of the material that was used to write the pattern IISER. The non-crystalline nature of the
50 nanotubes is clear from the SAED (Select Area Electron Diffraction) pattern that has been shown as an inset in Figure 4f.

**Factors controlling pattern formation**

We now describe in detail the factors which contribute to the
55 formation of the pattern, or more precisely the width of the pattern once the bubble is formed. These factors are as follows:

(i) Laser intensity and time of exposure of the laser on the sample: The laser intensity along with the duration for which the
60 laser is exposed to the sample, directly controls the size of the bubble formed. The size of the bubble in turn, determines the area over which material accumulation occurs and hence sets the width of the trail of deposited material. Such a difference in trail width for 26 mW and 66 mW laser power (focal spot size being
65 the same) are shown in Figure 5. For 26 mW power (Figure 5a), we observe a width of around 50 μm, which increased to around 70 μm at 66 mW (Figure 5b) – in both cases the laser exposure time was about 0.5 s. This demonstrates that the amount of material deposition is proportional to laser intensity for the same exposure time. On the other hand, Figures 2a and 2b show the increase in the size of the ring of material deposited on the glass slide when the same laser intensity was exposed on the sample for different exposure times (100 μs and 8 ms respectively). The actual values of the material ring size as a function of time have been plotted in Figure 2d – it can be seen that for 100 μs exposure time of the laser, the ring diameter is around 1 μm. The data fits well with a power law in time with the exponent value being about 0.30 ± 0.08, implying Ostwald ripening of the bubble.

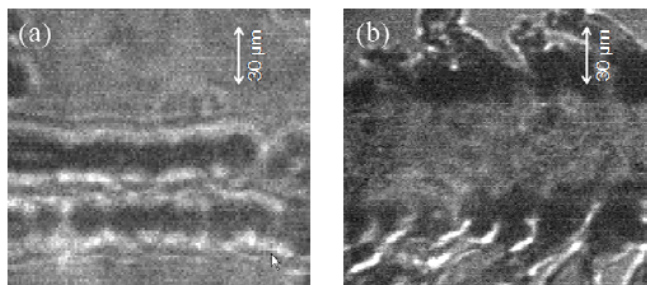

**Fig. 5** Controlling trail width with laser power: Optimal SOM based trails at different laser powers (a) 26 mW (b) 66 mW. The trail width is seen to vary proportionally with power. The trail width also depends upon the rate of moving the translation stage. In these sets of experiments, the stage was manually moved at constant speeds of around 1 mm/sec speeds to ensure a constant size of the bubble.

(ii) Speed of moving the translation stage: The speed of the stage essentially controls the time that the laser spends at a certain location, which determines the amount of material deposited on the substrate. Thus, to ensure uniform deposition of material and thereby uniform width of the trail, the stage is moved at a constant velocity. For instance, in the above experiments, the translation stage was moved manually at around 1 mm/s – the speed of translation being roughly calculated from the length of the trail divided by the total time taken to write it. Automation of the scanning process would further ensure uniformity in this regard.

(iii) Concentration of materials: The concentration of material in the solution controls the rate of material accretion from the dispersion at the site of the bubble. For example, for SOMs we observed bubble formation at around 0.1 M concentration (saturation concentration being around 0.3 M), whereas for the organic molecules, saturated solutions were used for trail formation. The reason was obviously the higher absorbance of SOMs compared to that of organic molecules.

## Conclusions

To summarize, we describe here a technique for completely controlled and fast (time scales less than 1 s in the present manual operation) optical patterning over a surface. A high density aqueous dispersion/solution of the material to be patterned is taken in a sample holder and exposed to an optical tweezers system. The high laser intensity at the focal point of the lens accompanied by absorption of the material at the laser wavelength leads to the formation of micro-bubbles that cause accretion of material from the liquid at the edges of the bubble due to Gibbs-Marangoni convection. The exposed region of the sample is then steered away from the laser spot using a scanning translation stage – the bubble follows the laser beam due to thermocapillary migration in a quasi-adiabatic manner so as to cause crystallization of the material in its wake on the overlying surface (top slide). Patterns can be formed of the deposited crystals by steering the stage in a pre-determined manner. The resolution for inscribing patterns can be controlled by the intensity of the laser in conjunction with the laser exposure time – presently, a resolution of around 1 μm has been achieved in depositing material over a glass substrate. We have patterned a host of materials including SOMs and organic molecules using this technique. SOMs have high absorbance at the wavelength of the trapping laser we use due to LMCT type transitions and can be patterned efficiently at much lower laser powers than has been used for patterning earlier.[26] In general any organic substrate with hydrogen bonding site/s or co-ordination site/s or with complementary charges as that of SOMs can be used for patterning/writing. However, since these have quite low absorbance in the IR, the preformed SOM has been used as a template to load them in order to exploit the LMCT absorption to minimize the intensity required of the laser. The loaded molecules can also additionally bring in novel functionality. For instance, fluorescent probes and potential drug moieties can be incorporated as functionalities on the material. The technique can have multiple and diverse applications ranging from printable electronics to controlled catalysis at the nano and micro scale. We are currently exploring some of these applications such as solution processed printable electronics[30-31] with water as well as other solvents in order to widen the scope of our technique.

## Acknowledgements

The authors gratefully acknowledge Dr. Subi George of JNCASR, Bangalore for providing perylene, and Dr. Anindita Bhadra of IISER Kolkata for preparing Figure 1. They also acknowledge the financial support of Indian Institute of Science Education & Research, Kolkata. S. R. further thanks DST Fast track for funding.

## Notes and references

[a] Department of Physical Sciences, Indian Institute of Science Education Research, Kolkata (IISER-Kolkata), Mohanpur 741252, India; E-mail: ayan@iiserkol.ac.in
[b] Department of Chemical Sciences, IISER-Kolkata, Mohanpur 741252, India; E-mail: cmreddy@iiserkol.ac.in;
[c] EFAML, Materials Science Centre, Department of Chemical Sciences, IISER-Kolkata, Mohanpur 741252, India; E-mail: s.roy@iiserkol.ac.in
[d] JNCASR, Bangalore, India.